\documentstyle[12pt]{article}
\def\as{\alpha_S}

\def\np#1#2#3{Nucl.\ Phys.\ B#1 (19#3) #2}
\def\pl#1#2#3{Phys.\ Lett.\ #1B (19#3) #2}

\def\zp#1#2#3{Zeit.\ Phys.\ C#1 (19#3) #2}

\begin{document}
\begin{titlepage}
\renewcommand{\thefootnote}{\fnsymbol{footnote}}
\begin{flushright}
     DFF 229/6/95\\   June 1995
     \end{flushright}
\par \vskip 10mm
\begin{center}
{\Large \bf Quarks and Gluons at Small $x$ \\
and Scaling Violation of $F_2$ \footnote{Talk given in the session on
{\it Proton Structure} at the Workshop on Deep Inelastic Scattering and QCD,
Paris, April 1995.}}
\end{center}
\par \vskip 2mm
\begin{center}
{\bf Stefano Catani}\\

\vskip 5 mm

{I.N.F.N., Sezione di Firenze}\\
{and Dipartimento
di Fisica, Universit\`a di Firenze}\\
{Largo E. Fermi 2, I-50125 Florence, Italy}
\end{center}

\par \vskip 2mm
\begin{center} {\large \bf Abstract} \end{center}
\begin{quote}
I present some comments on the
relationship between the small-$x$ behaviour of the parton (quark and gluon)
densities and the scaling violation of the proton structure function
$F_{2}(x,Q^{2})$.
 \end{quote}
\vspace*{\fill}
\begin{flushleft}
     DFF 229/6/95\\   June 1995
\end{flushleft}
\end{titlepage}
\renewcommand{\thefootnote}{\fnsymbol{footnote}}

\section{Introduction}

The main point I shall consider in this contribution (a more detailed
discussion can be found elsewhere~\cite{SDIS})
is whether the striking rise of
$F_{2}$, observed at
HERA~\cite{HERA}, calls for
a theoretical interpretation in terms of non-conventional QCD dynamics. Here,
non-conventional QCD
stands for any approach
(based either on the original BFKL equation~\cite{BFKL}
or on $k_{\perp}$-factorization~\cite{CCH,CH}) in which the small-$x$ behaviour
of $F_{2} (x,Q^{2})$ is studied by resumming logarithmic corrections of the
type $(\as \ln x)^{n}$ to {\em all orders} in the strong coupling $\as$.
By contrast, no small-$x$ resummation is performed within the conventional
QCD (or DGLAP~\cite{AP}) approach: the parton densities of
the proton at a fixed input scale $Q^{2}_{0}$ are evolved in $Q^{2}$ according
to the Altarelli-Parisi equation evaluated in {\em fixed-order}
perturbation theory.

It is certain that at asymptotically
small values of $x$, the fixed-order expansion in
$\as$ must become inadequate to describe the QCD
dynamics. However, since the DGLAP
approach successfully describes~\cite{S} the main features of HERA
data, the signal of non-conventional QCD dynamics (at least from
$F_{2}$, in the kinematic region explored at HERA so far) is {\em hidden}
or {\em mimicked} by a strong background of conventional QCD evolution.
\section{Scaling violation: the DGLAP approach}
The master equations for the small-$x$ behaviour of $F_2$ in perturbative QCD
are (symbolically) as follows~\cite{SDIS,S}
\begin{eqnarray}
F_{2} (x,Q^{2}) &\sim& {\tilde f}_{S} (x,Q^{2}) \;\;,\label{F2} \\
\frac{\partial F_{2} (x,Q^{2})}{\partial \ln Q^{2}} &\sim&
P_{SS}(\as(Q^{2}), x) \otimes {\tilde f}_{S}(x, Q^{2}) \nonumber \\
&+&
P_{Sg}(\as(Q^{2}), x) \otimes {\tilde f}_{g}(x,Q^{2}) \;\; ,
\label{dF2}
\end{eqnarray}
where $P_{ab}(\as,x)$ are the ({\em calculable})
splitting functions and ${\tilde f}_{S}$ and ${\tilde f}_{g}$ are the ({\em
phenomenological})
sea-quark and gluon densities.

The basis for Eqs.~(\ref{F2}),(\ref{dF2}) is provided by the the factorization
theorem of mass singularities. According to this theorem, splitting functions
and parton densities are not separately physical observables and, in
particular,
Eqs.~(\ref{F2}),(\ref{dF2}) refer to the so-called DIS scheme. However, when
evaluated in two-loop order, the splitting functions slightly depends on the
factorization scheme and the HERA data can be succesfully described~\cite{S}
by parton densities having the following small-$x$ behaviour,
$ {\tilde f}_{S} (x,Q^{2}_{0}) \sim x^{-\lambda_{S}}, \;
{\tilde f}_{g} (x,Q^{2}_{0}) \sim x^{-\lambda_{g}}$,
with
$\lambda_{S} = \lambda_{g} =
0.2 \div 0.3$ at the input scale $Q^{2}_{0} \sim$ 4 GeV${}^{2}$.
Actually, the HERA data may prefer~\cite{MRSG} $\lambda_{S} \neq \lambda_{g}$,
and, more precisely, $\lambda_{S} = 0.07 < \lambda_{g}
= 0.3 \div 0.35$.

Up to the second order in $\as$, the quark splitting functions $P_{SS}$ and
$P_{Sg}$ in Eq.~(\ref{dF2}) are essentially flat at small $x$. Thus,
the above results
tell us that the rise of $F_2$ at small $x$ is due to the DGLAP evolution in
the gluon channel combined with a steep behaviour $(\sim x^{-0.3})$ of the
input
densities at $Q^{2}_{0} \sim$ 4 GeV${}^{2}$. Moreover, taking seriously the
results of the MRS(G) analysis~\cite{MRSG}, one can argue that $F_2(x,Q^2)$
is {\em not} very
steep at $Q^{2}$-values of the order of few GeV${}^{2}$ (see Eq.~(\ref{F2})
with
$\lambda_S=0.07$) , but it is driven
by {\em strong} scaling violations (see Eq.~(\ref{dF2}) with $\lambda_g=0.35$).

Is there any room left for the non-conventional QCD approach? Is the power
behaviour of the input parton densities (independently of the actual values of
$\lambda_S$ and $\lambda_g$) related to small-$x$ resummation? Can we provide
an explanation for the (possibly) favoured values $\lambda_S < \lambda_g$?
\section{Scaling violation: small-$x$ resummation}
The above questions are formulated in the context of the parton picture and, as
recalled in Sec.~2, the parton densities have a well-defined physical meaning
only within the framework of the factorization theorem of mass singularities.
Therefore, in order to answer to these questions, we have to relate the
non-conventional QCD approach to the parton language.

A formalism which is able to combine {\em consistently}
small-$x$ resummation with the factorization theorem of
mass singularities has been set up in the last few years~\cite{CCH,CH}.
Within this formalism,
known as $k_{\perp}$-factorization or high-energy factorization,
one ends up with the usual QCD evolution
equations (\ref{F2}) and (\ref{dF2}), but the
splitting functions  $P_{ab}(\as,x)$ are no longer evaluated in fixed-order
perturbation theory. They are indeed supplemented with the all-order
resummation of the leading $(m=n-1)$, next-to-leading $(m=n-2)$ and,
possibly, subdominant $(m<n-2)$ contributions of the type
$\frac{1}{x} \as^{n} \ln^{m}x$ at small $x$.
More importantly, this resummation can be performed by having full control of
the factorization scheme dependence of splitting functions and parton
densities.

One of the main outcome of these studies is the calculation~\cite{CH} of the
quark splitting functions $P_{SS}$ and $P_{Sg}$ to next-to-leading
logarithmic accuracy in resummed perturbation theory. In particular,
these resummed splitting functions, evaluated {\em in the} DIS {\em scheme},
turn out to be much steeper than their two-loop expansions in perturbation
theory. Thus,
{\em stronger scaling violations} at small $x$,
were anticipated in Ref.~\cite{CH}.

Note that the {\em quark} splitting functions appear on the r.h.s. of the
master equation (\ref{dF2}).  The large value of $\partial F_{2}
(x,Q^{2})/\partial \ln Q^{2}$ measured at HERA
calls for a quite steep product
(convolution) $P_{Sg} \otimes {\tilde f}_{g}$.
In the DGLAP approach this condition  can be fulfilled only
by choosing a quite steep input distribution ${\tilde f}_{g}$.
However, after resummation, $P_{Sg}(\as,x)$ has a small-$x$ behaviour
which is much steeper than that in two-loop order.
Therefore, the use of resummed perturbation
theory at small $x$ may explain the scaling violations observed at
HERA without the necessity of
introducing a very steep input gluon density ${\tilde f}_{g}$. The results
of recent numerical analyses~\cite{EHW} support this conclusion.

There is also an alternative (and more striking) way to restate
the same conclusion on the possible relevance of small-$x$ resummation for
the HERA data on $F_{2}$. So far, I have only discussed the DIS
scheme.
One can consider a different factorization
scheme, {\em the} SDIS {\em scheme}~\cite{SDIS},
in which the
resummation effects discussed above are removed from the quark splitting
functions and absorbed into the redefinition of the {\em gluon} density.
In the new scheme, $i)$~the resummed quark splitting functions
differ slightly from the corresponding two-loop functions in the DIS scheme
and $ii)$ a steep
gluon density (in particular, a gluon density steeper than the quark
density) arises naturally as the result of small-$x$ resummation. From
the property $i)$, it follows that the analysis of the scaling violations of
$F_2$ in the SDIS scheme
is very similar to that in the DGLAP approach. Thus, the property $ii)$
offers a qualitative explanation of the results in Ref.~\cite{MRSG}:
the MRS(G) partons with $\lambda_g > \lambda_S$ may be interpreted
as the partons in the resummed SDIS scheme.
\section{Conclusion}
In summary, HERA may have seen a (weak) signal of non-conventional small-$x$
dynamics not in the (absolute) steep rise of $F_2$ but {\em rather} in stronger
scaling
violations at moderate values of $Q^2$. More definite conclusions demand
further
phenomenological investigations
and more accurate
data on $F_2$ in a range of $x$ and $Q^2$ as largest as possible.
\begin{center}
{\large\bf Aknowledgements}
\end{center}
This research is supported in part by
EEC Programme {\it Human Capital and Mobility}, Network {\it Physics at High
Energy Colliders}, contract CHRX-CT93-0357 (DG 12 COMA).

%

\end{document}